\documentstyle[twoside,psfig]{article}

\catcode`\@=11
\long\def\@makefntext#1{
\protect\noindent \hbox to 3.2pt {\hskip-.9pt  
$^{{\eightrm\@thefnmark}}$\hfil}#1\hfill}		

\def\@makefnmark{\hbox to 0pt{$^{\@thefnmark}$\hss}}	
	
\def\ps@myheadings{\let\@mkboth\@gobbletwo
\def\@oddhead{\hbox{}
\rightmark\hfil\eightrm\thepage}   
\def\@oddfoot{}\def\@evenhead{\eightrm\thepage\hfil
\leftmark\hbox{}}\def\@evenfoot{}
\def\sectionmark##1{}\def\subsectionmark##1{}}

\oddsidemargin=\evensidemargin
\newcounter{sectionc}\newcounter{subsectionc}\newcounter{subsubsectionc}
\renewcommand{\section}[1] {\vspace{12pt}\addtocounter{sectionc}{1} 
\setcounter{subsectionc}{0}\setcounter{subsubsectionc}{0}\noindent 
	{\tenbf\thesectionc. #1}\par\vspace{5pt}}
\renewcommand{\subsection}[1] {\vspace{12pt}\addtocounter{subsectionc}{1} 
	\setcounter{subsubsectionc}{0}\noindent 
	{\bf\thesectionc.\thesubsectionc. {\kern1pt \bfit #1}}\par\vspace{5pt}}
\renewcommand{\subsubsection}[1] {\vspace{12pt}\addtocounter{subsubsectionc}{1}
	\noindent{\tenrm\thesectionc.\thesubsectionc.\thesubsubsectionc.
	{\kern1pt \tenit #1}}\par\vspace{5pt}}
\newcommand{\nonumsection}[1] {\vspace{12pt}\noindent{\tenbf #1}
	\par\vspace{5pt}}

\newcounter{appendixc}
\newcounter{subappendixc}[appendixc]
\newcounter{subsubappendixc}[subappendixc]
\renewcommand{\thesubappendixc}{\Alph{appendixc}.\arabic{subappendixc}}
\renewcommand{\thesubsubappendixc}
	{\Alph{appendixc}.\arabic{subappendixc}.\arabic{subsubappendixc}}

\renewcommand{\appendix}[1] {\vspace{12pt}
        \refstepcounter{appendixc}
        \setcounter{figure}{0}
        \setcounter{table}{0}
        \setcounter{lemma}{0}
        \setcounter{theorem}{0}
        \setcounter{corollary}{0}
        \setcounter{definition}{0}
        \setcounter{equation}{0}
        \renewcommand{\thefigure}{\Alph{appendixc}.\arabic{figure}}
        \renewcommand{\thetable}{\Alph{appendixc}.\arabic{table}}
        \renewcommand{\theappendixc}{\Alph{appendixc}}
        \renewcommand{\thelemma}{\Alph{appendixc}.\arabic{lemma}}
        \renewcommand{\thetheorem}{\Alph{appendixc}.\arabic{theorem}}
        \renewcommand{\thedefinition}{\Alph{appendixc}.\arabic{definition}}
        \renewcommand{\thecorollary}{\Alph{appendixc}.\arabic{corollary}}
        \renewcommand{\theequation}{\Alph{appendixc}.\arabic{equation}}
        \noindent{\tenbf Appendix \theappendixc #1}\par\vspace{5pt}}
\newcommand{\subappendix}[1] {\vspace{12pt}
        \refstepcounter{subappendixc}
        \noindent{\bf Appendix \thesubappendixc. {\kern1pt \bfit #1}}
	\par\vspace{5pt}}
\newcommand{\subsubappendix}[1] {\vspace{12pt}
        \refstepcounter{subsubappendixc}
        \noindent{\rm Appendix \thesubsubappendixc. {\kern1pt \tenit #1}}
	\par\vspace{5pt}}

\newcommand{\textlineskip}{\baselineskip=13pt}
\newcommand{\smalllineskip}{\baselineskip=10pt}

\def\eightcirc{
\begin{picture}(0,0)
\put(4.4,1.8){\circle{6.5}}
\end{picture}}
\def\eightcopyright{\eightcirc\kern2.7pt\hbox{\eightrm c}} 

\newcommand{\copyrightheading}[1]
	{\vspace*{-2.5cm}\smalllineskip{\flushleft
	{\footnotesize International Journal of Modern Physics D #1}\\
	{\footnotesize $\eightcopyright$\, World Scientific Publishing
	 Company}\\
	 }}

\newcommand{\publisher}[2]{{\begin{center}\footnotesize\smalllineskip 
	Received #1\\
	Revised #2
	\end{center}
	}}

\def\abstracts#1#2#3{{
	\centering{\begin{minipage}{4.5in}\footnotesize\baselineskip=10pt
	\parindent=0pt #1\par 
	\parindent=15pt #2\par
	\parindent=15pt #3
	\end{minipage}}\par}} 

\newcommand{\bibit}{\nineit}

\renewenvironment{thebibliography}[1]
	{\frenchspacing
	 \ninerm\baselineskip=11pt
	 \begin{list}{\arabic{enumi}.}
	{\usecounter{enumi}\setlength{\parsep}{0pt}
	 \setlength{\leftmargin 12.7pt}{\rightmargin 0pt} 
	 \setlength{\itemsep}{0pt} \settowidth
	{\labelwidth}{#1.}\sloppy}}{\end{list}}

\newcounter{itemlistc}
\newcounter{romanlistc}
\newcounter{alphlistc}
\newcounter{arabiclistc}

\newcommand{\fcaption}[1]{
        \refstepcounter{figure}
        \setbox\@tempboxa = \hbox{\footnotesize Fig.~\thefigure. #1}
        \ifdim \wd\@tempboxa > 5in
           {\begin{center}
        \parbox{5in}{\footnotesize\smalllineskip Fig.~\thefigure. #1}
            \end{center}}
        \else
             {\begin{center}
             {\footnotesize Fig.~\thefigure. #1}
              \end{center}}
        \fi}

\newcommand{\tcaption}[1]{
        \refstepcounter{table}
        \setbox\@tempboxa = \hbox{\footnotesize Table~\thetable. #1}
        \ifdim \wd\@tempboxa > 5in
           {\begin{center}
        \parbox{5in}{\footnotesize\smalllineskip Table~\thetable. #1}
            \end{center}}
        \else
             {\begin{center}
             {\footnotesize Table~\thetable. #1}
              \end{center}}
        \fi}

\def\@citex[#1]#2{\if@filesw\immediate\write\@auxout
	{\string\citation{#2}}\fi
\def\@citea{}\@cite{\@for\@citeb:=#2\do
	{\@citea\def\@citea{,}\@ifundefined
	{b@\@citeb}{{\bf ?}\@warning
	{Citation `\@citeb' on page \thepage \space undefined}}
	{\csname b@\@citeb\endcsname}}}{#1}}

\newif\if@cghi
\def\cite{\@cghitrue\@ifnextchar [{\@tempswatrue
	\@citex}{\@tempswafalse\@citex[]}}
\def\citelow{\@cghifalse\@ifnextchar [{\@tempswatrue
	\@citex}{\@tempswafalse\@citex[]}}
\def\@cite#1#2{{$\null^{#1}$\if@tempswa\typeout
	{IJCGA warning: optional citation argument 
	ignored: `#2'} \fi}}

\def\pmb#1{\setbox0=\hbox{#1}
	\kern-.025em\copy0\kern-\wd0
	\kern.05em\copy0\kern-\wd0
	\kern-.025em\raise.0433em\box0}

\def\fnt#1#2{\footnotetext{\kern-.3em
	{$^{\mbox{\scriptsize #1}}$}{#2}}}

\def\@makefnmark{\hbox to 0pt{$^{\@thefnmark}$\hss}}	
	
\def\ps@myheadings{%
    \let\@oddfoot\@empty\let\@evenfoot\@empty
    \def\@evenhead{\slshape\leftmark\hfil}
    \def\@oddhead{\hfil{\slshape\rightmark}}
    \let\@mkboth\@gobbletwo
    \let\sectionmark\@gobble
    \let\subsectionmark\@gobble
    }
\font\tenrm=cmr10
\font\tenit=cmti10 
\font\tenbf=cmbx10
\font\bfit=cmbxti10 at 10pt
\font\ninerm=cmr9
\font\nineit=cmti9

\font\eightrm=cmr8

\def\qed{\hbox{${\vcenter{\vbox{			
   \hrule height 0.4pt\hbox{\vrule width 0.4pt height 6pt
   \kern5pt\vrule width 0.4pt}\hrule height 0.4pt}}}$}}

\begin{document}
\setlength{\textheight}{7.7truein}  

\thispagestyle{empty}

\markboth{\protect{\footnotesize\it Quark-diquark matter equation of 
state $\ldots$}}{\protect{\footnotesize\it  Quark-diquark matter equation of 
state  $\ldots$}}

\normalsize\textlineskip

\setcounter{page}{1}

\copyrightheading{}	

\vspace*{0.88truein}

\centerline{\bf QUARK-DIQUARK MATTER EQUATION OF STATE}
\vspace*{0.035truein}
\centerline{\bf AND COMPACT STAR STRUCTURE}
\vspace*{0.37truein}

\centerline{\footnotesize G. LUGONES\footnote{Email: glugones@iagusp.usp.br}   and J. E. HORVATH}
\baselineskip=12pt
\centerline{\footnotesize\it Instituto Astron\^omico e Geof\'{\i}sico, Universidade de S\~ao Paulo,}
\baselineskip=10pt
\centerline{\footnotesize\it Av. M. St\'efano 4200, Agua Funda, 04301-904 S\~ao Paulo SP, Brazil}
\vspace*{10pt}
\vspace*{0.225truein}
\publisher{(received date)}{(revised date)}

\vspace*{0.21truein}
\abstracts{We present the equation of state (EOS) of quark-diquark matter in the quark 
mass-density-dependent model. 
The region of the $2-D$ parameter space 
inside which this quark-diquark matter is stable against diquark 
disassembling and 
hadronization is determined. 
Motivated by observational data suggesting a high 
compactness of some neutron stars (NS) we present models based on the 
present EOS and compare the results with those obtained with previous 
works addressing a quark-diquark composition based on the MIT Bag model. 
We show that very compact self-bound stars (yet having $\geq 1 M_{\odot}$) 
are allowed by our EOS even if 
the diquark itself is unbound.}{}{}

\vspace*{1pt}\textlineskip	
\section{Introduction}	
\vspace*{-0.5pt}
\noindent
The spectacular advances of astronomical instrumentation in the last decade has 
stimulated a great deal of activity on "old" questions of relativistic 
astrophysics. Among the latter we find the internal structure of compact 
stars, 
which is not only interesting {\it per se}, but also important for a through 
knowledge of the QCD diagram in the low-$T$ , high $\mu$ region. 

One of the most simple forms of (indirectly) investigating the nature of 
high-density matter, more precisely of the EOS is by obtaining accurate 
determinations of 
masses and radii. In fact, leaving aside the possibility of differences due 
to extreme rotation, and adopting General Relativity as a framework, a 
comparison 
of the static models generated by integration of the 
Tolman-Oppenheimer-Volkoff 
equations with observed data should reveal a great deal of information about 
the equation of state. Only recently we have achieved sufficient accuracy to 
attempt some key tests 
on selected objects, although it should be acknowledged that three decades 
of 
compact star astrophysics had produced quite clever arguments to determine 
masses and radii, even in the cases in which they proved to be wrong. 

A paradigmatic example of the above assertions is the celebrated 
determination 
of the binary pulsar mass PSR 1913+16, believed to be accurate to the fourth 
decimal place \cite{vankerkwijk}. Other methods based on combinations of 
spectroscopic and 
photometric techniques have been recently perfectioned, and have confirmed 
that at least one X-ray source is significantly above the "canonical" 
1.44 $M_{\odot}$; namely Vela X-1 for which a value of 
$1.87^{+0.23}_{-0.17} \, M_{\odot}$ has been obtained \cite{vankerkwijk}. 
Some works on compact stars have attracted the attention recently 
because not only the masses have been determined, but also indications of 
the radii are available, suggesting a very compact structure. 

On the other hand, from a particle physics point of view 
it is fairly well established that at high temperatures and/or high 
baryon number densities confinement is effectively disabled. The old 
question of 
whether the deconfined matter takes place inside "neutron stars" is not yet 
settled, although a great deal of interest in this issue emerged from recent 
calculation of superconducting QCD phases (two-flavor superconducting (2SC) 
and "color flavor locking" (CFL), 
see \cite{Al} for a review). 
In fact, there is growing evidence that paired states or {\it diquarks} (e. 
g. a spin-0, color-antitriplet 
bound state of two quarks) might occur as a component in the QCD plasma. 
This is important 
for understanding hadron structure and high energy particle 
reactions. Of course, at extremely high densities diquarks are expected to 
lose their 
identity and dissolve into quarks. In the intermediate phase diquarks would 
be expected 
to be favored. 

Diquark correlations were originally thought to 
arise in part from spin-dependent interactions between two quarks. Later the 
non-perturbative interactions were claimed to produce a gap substantially 
greater than in the '80s, possibly as large as $\sim 100 \, MeV$. 
Although there is no consensus about diquark details, it seems certain that 
a $(ud)$ quark pair 
experiences some attraction when in total spin-0 and colour-3* states. Many 
works assume that the diquark correlation is similar to the one experienced by  electrons 
in a superconductor, with diquarks being the Cooper pairs of QCD 
\cite{Fredriksson}. 
However, we shall not be concerned with the details of the 
binding of the diquark itself in this work. This is because for compact star astrophysics 
it is perhaps more important to look for the full 
free energy (per baryon) behavior in bulk rather than the 
mass of the diquark individually. 

\section{The equation of state} 
\noindent
The quark mass-density-dependent model (QMDD model) is a 
phenomenological description that attempts to incorporate the confinement 
and asymptotic 
freedom of quarks as a function of the baryonic density of matter. 
According to lattice calculations \cite{belyaev} and string model 
investigations \cite{isgur}, 
the quark-quark interaction is proportional to the distance so that 
$m \propto n_B^{-1/3}$ may be assumed \cite{peng2000c61}. Therefore, it is reasonable to 
adopt the 
following parametrizations of the particle masses. For the quarks $u$ and 
$d$ 
we define as usual 
\cite{benvenuto95,lugones95,peng97,peng99,peng2000c61,peng2000c62} 

\begin{equation} 
m_u = m_d = \frac{C}{ n_B^{1/3} }. 
\end{equation} 

For diquarks, we introduce here the mass parameter $m_{D0}$ and write the 
diquark mass $m_{D}$ as 

\begin{equation} 
m_D = m_{D0} + m_u + m_d = m_{D0} + \frac{2 C}{ n_B^{1/3} }. 
\label{massmd} 
\end{equation}

The non-perturbative character of the interactions between quarks and 
diquarks is included through this 
particular mass parametrization with the baryon number density. 

Before describing the equation of state in detail we would like to make a 
number of qualitative remarks: 
To make contact with current calculations we must determine to which extent 
the model describes the expected diquark physics. The diquark pairing gap 
that arises from current calculations \cite{Al} clearly depends on the chemical 
potential (or on the baryon number density). 
Qualitatively the gap is zero at low densities, reaches a maximum value in 
the 10-100 MeV range at intermediate 
densities and goes to zero at high densities. To mimic this behaviour the 
parameter $m_{D0}$ 
should have the same shape (but with negative sign) as a function of, for 
example, the baryon number density, even though in QMDD models it is 
introduced to simulate confinement only. We shall consider only constant 
values for $m_{D0}$ in this work. 
As we shall see below this approximation is a reasonable starting point for 
the physics of interest to us here. 
On the other hand we must note that recent lattice simulations show that the 
diquark could be unbound \cite{Biel} 
The latter situation can be described by the present model with positive 
values of $m_{D0}$, which are also included in the analysis (see below). 
We must remark that for negative values of $m_{D0}$ the diquark mass becomes 
negative (as expected) for some 
baryon number density beyond which the model is no longer valid. However, 
this $n_B$ is very 
high for typical values of $m_{D0}$ (for example, $m_{D0}$ = -100 MeV gives 
$n_{B,lim}$ greater than $12 \times n_0$, with $n_0 = 0.16$ fm$^{-3}$ the 
nuclear saturation density. 
We note that our quark-diquark phase may be directly associated to the phase 
explicitely addressed by  Rapp et al. \cite{Rapp} and termed QDQ, as a state intermediate 
between hadrons and 2SC, but an identification with the 2SC may also be 
possible when the masses go to zero at high density.
Although strictly speaking this QDQ and 2SC with zero net strangeness would 
be ustable against  weak interaction decays in neutron stars, 
the number of strange quarks that is produced in the mixture can be 
neglected in a first approximation. 
The finite mass $m_s \sim 150$ MeV suppress the ocupation of the 
corresponding Fermi sea, while 
the occupation of the bosonic QDQ or 2SC ground state is strongly favoured 
energetically. 
We must note that a similar mass parametrization ( $m_s = m_{s0} + C/ 
n_B^{1/3}$) has been employed 
to describe strange quark matter within the QMDD model. Also note the 
slightly different 
parametrization than the one employed in Ref. \cite{HLP} motivated by an 
intuitive construction of the mass formula. 

In view of the above, we shall proceed to model the quark - diquark plasma 
as a gas 
of $u$ and $d$ quarks, and $(ud)$ diquarks. Non-perturbative 
effects are accounted for using the quark mass-density-dependent model of 
confinement to derive the properties of the equation of state. 

The expression for the thermodynamic potential density of particles of mass 
$m_i$, 
chemical potential $\mu_i$ and degeneracy factor $g_i$ in the large 
volume limit of a free gas is 

\begin{equation} 
\Omega_i = \frac{g_i T}{2 \pi^2} \int_0^{\infty} 
\ln (1 \pm e^{(\mu_i - (p^2 + m_i^2)^{1/2} )/T)} ) p^2 dp 
\label{omega1} 
\end{equation} 

Because of the dependence of the mass with density $n_B$ in the QMDD model 
the standard relation $P= -\Omega$ is no longer valid.
\cite{peng2000c61,peng2000c62,benvenuto95,lugones95}  
The pressure is actually given by 

\begin{equation} 
P = -\frac{ \partial (V \Omega)}{\partial V } = 
\sum_i \bigg( - \Omega_i + n_B \frac{\partial m_i}{\partial n_B} 
\frac{\partial \Omega_i}{\partial m_i} 
\bigg) . 
\label{p} 
\end{equation} 

For the other thermodynamical quantities we write (see Ref.[12]) 

\begin{equation} 
E = \Omega + \sum_i \mu_i n_i - T \frac{\partial \Omega}{\partial T} 
\label{E} 
\end{equation} 

\begin{equation} 
n_i = - \frac{\partial \Omega}{\partial \mu_i} \bigg|_{T,m_k} 
\label{n2} 
\end{equation} 

\noindent with $E$ the energy density and $n_i$ the particle number density 
of the $i$-species. 

As it has been stressed before \cite{benvenuto95,lugones95}, for the 
fermionic component the 
term $n_B \frac{\partial m_i}{\partial n_B} \frac{\partial 
\Omega_i}{\partial m_i}$ forces the pressure to be zero at finite 
baryon number, playing the role of the bag constant $B$ of the MIT 
Bag Model. Needless to say, this is a property related to the mass 
parametrization only, and has no relation with the 
statistics of the particles themselves (Bose or Fermi). 
However, it happens that in the case of bosons this extra term does not 
contribute at zero temperature. 
Thus, as in the standard case, for massive bosons at $T=0$ the energy 
density reduces to $E = \sum_i n_i m_i$. 

As we mentioned above we shall model the quark - diquark plasma at zero 
temperature as a mixture of a Fermi relativistic gas 
of $u$ and $d$ quarks, and a Bose condensate of $(ud)$ diquarks. 
We neglect the contribution of antiparticles and that of a leptonic 
component such as 
electrons, muons or neutrinos. 
This mixture is then described by \cite{shapiro,chandra} 

\begin{equation} 
P = \sum_{i=u,d} m_i n_i x_i^2 G(x_i) - \sum_{i=u,d} m_i n_i f (x_i) 
\end{equation} 

Diquarks do not contribute to the pressure in this approximation, since they 
are all in the ground state, but do contribute to the energy density 

\begin{equation} 
E = \sum_{i=u,d} m_i n_i F(x_i) + m_D n_D . 
\end{equation} 

The number density of quarks $u$ and $d$ is given by 

\begin{equation} 
n_i = \frac{1}{\pi^2} m_i^3 x_i^3 . 
\end{equation} 

where the functions $F, G$ and $f$ are given by 
\cite{benvenuto95,peng2000c62} 

\begin{equation} 
F(x_i) = \frac{3}{8} \bigg[ x_i (x_i^2 + 1)^{1/2} (2x_i^2 + 1 ) - 
\arg \sinh(x_i) \bigg] / x_i^3 
\end{equation} 

\begin{equation} 
G(x_i) = \frac{1}{8} \bigg[ x_i (x_i^2 + 1)^{1/2} (2x_i^2 - 3 ) + 3 
\arg \sinh(x_i) \bigg] / x_i^5 
\end{equation} 

\begin{equation} 
f(x_i) = -\frac{3}{2} \frac{n_B}{m_i} \frac{dm_i}{dn_B} \bigg[ x_i 
(x_i^2 + 1)^{1/2} 
- \arg \sinh(x_i) \bigg] / x_i^3 
\end{equation} 

\noindent being $x = p_{F} / m_i = (\mu_i^2 - m _i^2)^{1/2} / m_i $, $\mu_i$ 
the chemical potential 
and $m_i$ the mass of the particle, which in our case depends on the baryon 
density.

These equations must be complemented with the following conditions: 

1) Chemical equilibrium: $D \rightleftharpoons u + d$ requires $\mu_D = 
\mu_u + \mu_d$. 
If we use the fact that at $T=0$ the condensed diquarks all have 
zero kinetic energy we have ($\mu_D = m_D$) 

\begin{equation} 
m_D = \mu_u + \mu_d . 
\end{equation} 

Then, using the fact that, $\mu = (x^2 +1)^{1/2} m$, $m_u = m_d$, and 
$m_D = m_{D0} + 2 m_u$, we can write the chemical equilibrium as 

\begin{equation} 
\frac{m_{D0}}{m_u} + 2 = (x_{u}^{2} +1)^{1/2} + (x_{d}^{2} +1)^{1/2} . 
\label{chem_eq} 
\end{equation}

2) Electrical charge neutrality implies $ 1/3 n_D + 2/3 n_u - 1/3 n_d = 
0$, therefore, 
the diquark number density is given by 

\begin{equation} 
n_D = n_d - 2 n_u = \frac{1}{\pi^2} m_u^3 (x_d^3 - 2 x_u^3 ) . 
\end{equation}

3) Baryon number is given by $n_B = 2/3 n_D + 1/3 (n_u + n_d)$, thus 

\begin{equation} 
n_B = n_d - n_u = \frac{1}{\pi^2} m_u^3 (x_d^3 - x_u^3 ). 
\end{equation}

The equation of state can be solved numerically for each $n_B$ once the 
parameters $(C, m_{D0})$ are given (see Figs 1 and 2).

As the baryon density increases the EOS tends to a pure "quark d-diquark" 
composition since the number density of quarks $u$ goes to zero and the 
abundances of quarks $d$ and diquarks $D$ become equal. The same behaviour 
arises for smaller enough values of the mass $m_{D0}$ (see below). In these 
cases the EOS 
takes a very simple form: $x_u = 0$, $n_u = 0$, $n_B = n_D = n_d = ( m_u^3 
x_d^3 ) / \pi^2$.

As expected from simple physical considerations, 
we find that the EOS of the quark-diquark is quite soft, especially at relatively high densities, 
because of diquark condensation. The EOS becomes somewhat stiffer at low 
densities when compared to the MIT-inspired EOS developed in Ref.[18], at least for a large 
region of the parameter space. 
The same effect is found if we compare the MIT and the 
QMDD model EOS for strange quark matter \cite{benvenuto95,lugones95}.

\section{Stability of the quark - diquark matter} 
\noindent
In order to determine the allowed values for $C$ and $m_{D0}$ we impose the 
following stability 
criteria for quark-diquark matter. 

1) Instability of two flavor quark matter: The energy per baryon of two 
flavor quark matter (at $P=0$ and $T=0$) must be higher than 939 MeV (the 
neutron mass). Assuming that the current mass of quarks $u$ and $d$ is zero 
this results in the condition $C > (159.3 \,  MeV)^2$. 

The energy per baryon of pure $udd$ matter at $P=0$ and $T=0$ is, for a 
given $C$ 

\begin{equation} 
\frac{E}{n_B} \bigg|_{udd} = m^{'}_u [ F(x_{u0}) + 2 F(x_{d0}) ] 
\label{enb_udd} 
\end{equation}

\noindent where $m^{'}_u$ is the mass of quark $u$ in two flavor quark 
matter at zero $P$ and $T$, 

\begin{equation} 
m^{'}_u = \frac{C}{n_B^{1/3}} = \frac{C}{n_u^{1/3}} = 
\frac{C^{1/2} \pi^{1/3}}{x_{u0}^{1/2}} 
\end{equation} 

\noindent and $x_{d0} = 2^{1/3} x_{u0}$, where $ x_{u0} = 1.13173096$ is 
the value that produces a zero-point pressure, $P_{udd} = 0$. 
This criterion has already been used in previous papers 
\cite{benvenuto95,lugones95,peng2000c61,peng2000c62}.

2) Relative stability of quark-diquark matter: The energy per baryon of 
quark - diquark matter 
(hereafter ud-D matter) must be lower than 
that of pure two flavor quark matter (at $P=0$, $T=0$). 

The zero pressure condition for quark - diquark matter at $T=0$ reads 

\begin{equation} 
x_u^5 G(x_u) - x_u^3 f(x_u) + x_d^5 G(x_d) - x_d^3 f(x_d) = 0 . 
\label{P_nulo} 
\end{equation} 

The energy per baryon is given by 

\begin{equation} 
\frac{E}{n_B} \bigg|_{ud-D} = m_u \frac{x_u^3 F(x_u) + x_d^3 F(x_d) + 
(\frac{m_{D0}}{m_u}+2) (x_d^3 - 2 x_u^3) }{x_d^3 - x_u^3} 
\label{enb_diq} 
\end{equation} 

\noindent 
Note that the mass $m_u$ of quark $u$ in $ud-D$ matter is in fact a 
function of $C$, $x_u$ and $x_d$ given by $m_u = C / n_B^{1/3} = C / (n_d - 
n_u)^{1/3}$, from which we derive 

\begin{equation} 
m_u = \frac{ C^{1/2} \pi^{1/3} }{(x_d^3 - x_u^3)^{1/6}} . 
\end{equation} 

\noindent 
Also, the factor $(m_{D0}/m_u +2)$ in Eq. (\ref{enb_diq}) is a function of 
$x_u$ and $x_d$ given by Eq. (\ref{chem_eq}).

The fundamental relation for this stability requirement is 

\begin{equation} 
\frac{E}{n_B} \bigg|_{ud-D} < \frac{E}{n_B} \bigg|_{udd} 
\label{igual1} 
\end{equation} 

\noindent 
therefore, from eqs. (\ref{enb_diq}) and (\ref{enb_udd}) we obtain 

\begin{equation} 
\frac{x_u^3 F(x_u) + x_d^3 F(x_d) + (\frac{m_{D0}}{m_u}+2) (x_d^3 - 2 
x_u^3) }{(x_d^3 - x_u^3)^{7/6}} 
< \kappa_0 
\label{igual2} 
\end{equation} 

\noindent 
where 

\begin{equation} 
\kappa_0 = \frac{ [ F(x_{u0}) + 2 F(x_{d0}) ] }{x_{u0}^{1/2}} . 
\label{igual22} 
\end{equation} 

From eqs. (\ref{igual2}) and (\ref{P_nulo}) we find numerically the set 
of values $x_{u1}$ and $x_{d1}$ that satisfy 
the imposed stability conditions. 

Using chemical equilibrium alone (eq. (\ref{chem_eq})) we find a very simple 
relation between 
$m_{D0}$ and $C$ 

\begin{equation} 
m_{D0} = K C^{1/2} 
\end{equation} 

\noindent 
with 

\begin{equation} 
K = \pi^{1/3} \frac{(x_{u1}^2 + 1)^{1/2} + (x_{d1}^2 + 1)^{1/2} - 
2}{(x_{d1}^3 - x_{u1}^3)^{1/6} } . 
\end{equation} 

This is an "isochemical" straight line along which the energy per baryon 
increases with $C$. 
The condition $ (E/ n_B ) |_{ud-D} < (E / n_B) |_{udd}$ 
is fullfilled for $K < 1.723$ (the upper limit is given by the full straight 
line in Figure 3). Also, we find that for $K < 0.8417$ 
there are no quarks $u$ in the mixture and the composition is always a pure 
"quark d-diquark" mixture (see dashed line in Figure 3). 

3) Absolute stability condition: we have also looked for a region inside 
which quark - diquark matter could have even a lower energy per baryon than 
the neutron mass 

\begin{equation} 
\frac{E}{n_B} \bigg|_{ud-D} < 939 MeV, 
\label{igual1} 
\end{equation} 

\noindent 
being though absolutely stable. In the $(C^{1/2}, m_{D0})$ parameter space 
this region is 
"triangle-like". The same shape has been obtained for strange quark matter 
in the QMDD model \cite{lugones95}. 
On the right side boundary of this "triangle-like" region the energy per 
baryon of quark-diquark matter at zero pressure 
is exactly 939 MeV. The region is shown in Fig. 3 and labeled with $A$ and 
$A^{'}$. 
It is interesting to note that negative values of 
$m_{D0}$ are allowed (see Fig. 3) and so the diquark may be bound in 
this model, that is, its mass will be smaller than the sum of the masses of its constituent 
particles. 

\section{Quark-diquark stars} 
\noindent
The exploration of the mass-radius relation for compact stars is in fact a 
potentially powerful tool in testing the existence of different phases of matter inside 
NSs, and their composition. On the other hand the observational data 
of the mass-radius relation 
imposes severe constaints on high density equations of state. 

Almost all of the meassured masses of "neutron stars" lie within a narrow 
range around 1.4 $M_{\odot}$. 
This mass scale may be due to the fact that neutron stars are formed in the 
gravitational collapse 
of supernovae and come just from the iron cores of core collapse supernovae. 
However, we may wonder whether smaller neutron stars could exist and whether 
there exist some mechanism 
capable of create them because these are not expected in core collapse 
models and hold the potential for discriminating among possible equations of 
state. This question becomes more significant in view of the recent 
claim of very 
low-mass/low-radii objects. At least two objects, Her X-1 and RX J1856-37 
are candidates to 
high compactness. They have been claimed to have radii $\sim 7 km$ and a 
masses around $1 M_{\odot}$ \cite{li,Pons}.

If quark- diquark matter has a lower energy per baryon than normal nuclear 
matter we may expect the existence of stable stars made up entirely by this phase. 
By solving the Tolman-Oppenheimer-Volkoff equations of stellar structure, 
we explored the mass- radius relation for quark-diquark stars assuming 
a parameter set inside the window of absolute stability.

In Figs. 4 and 5  we see that for particular values of the parameters $(C^{1/2}, m_{D0})$ 
the obtained values of $M$ and $R$ are compatible with present determinations of 
the above sources. 
We also find that the effect of increasing $m_{D0}$ for a fixed $C$ is to lower both 
the maximum radius and the maximum mass of the stellar configurations. 
Similarly, increasing $C$ at fixed $m_{D0}$ allows configurations with 
smaller maximum radius and mass. 
The $(C^{1/2}, m_{D0})$ values selected for the sequences shown in Figs. 4 and 5
match naturally the observed compactness of Her X-1 and RX J1856-37.

\section{Conclusions} 
\noindent
We derived an equation of state for quark diquark-matter in a 
self-consistent phenomenological model where the mass of quarks and 
diquarks 
is parametrized with the baryon number density. The confining forces are 
included via a dependence of the mass with 
$n_B^{-1/3}$ . The model depends on two free parameters $(C^{1/2}, m_{D0})$ 
whose values can be 
limited invoking stability criteria and allows the existence of a stable 
state of the 
quark-diquark gas at zero pressure, for densities of the order of few times 
the nuclear saturation density (see Figs 1 and 2). 
The free parameter $m_{D0}$ is introduced in order to describe in a simple 
form the unknown binding 
of the diquark. Negative values of $m_{D0}$ should mimic the behaviour of 
bound color superconducting 
quark matter while positive values are considered in view of recent results \cite{Biel}
claiming a diquark mass of $\sim 700$ 
MeV  which suggest that this particle is unbound. 
The mixture contains a Bose condensate of $(ud)$ diquarks so that, the 
resulting equation of state is 
in fact very soft, as it can be seen in Figure 1 where we have depicted 
$P$ versus $E$ for different parameters . 
The here described quark-diquark phase has always 
a non-zero fermionic pressure, since due to the excess of quarks $d$ over 
quarks $u$ there will always exist a 
free Fermi gas of quarks $d$ to satisfy the equilibrium conditions. 
When compared with the equation of state for quark-diquark 
matter presented in \cite{horvath1} we found  that independently of the value of 
the parameters, the latter  is considerably stiffer 
at low densities. 
However, in some cases (c,d, e and f of Fig. 1) there is a limiting density beyond which the 
new equations of state are softer than the previous one. This is relevant for compact star 
structure 
because it allows the existence of stable compact stars with even smaller 
radius than those calculated in previous models \cite{horvath1}. It has been argued 
that the few percent 
effects ($\cal{O}$ $(\Delta  /\mu)^{2}$) from the condensation are impossible to 
disentangle from other uncertainties, but the fact is that, for example,  the whole strange 
matter theory is also subject to this caveat, and in this respect the models 
here presented open a qualitatively similar scenario for compact stars.

We have found the regions in the parameter space $(C^{1/2}, m_{D0})$ where 
quark-diquark matter in bulk is stable against 
decaying to pure two flavor quark matter and to a neutron gas (see Fig. 3). 
The region of absolute stability allows the existence of pure quark-diquark 
stars. 
It is important to note that the existence of self-bound compact stars is 
possible 
for both bound and $unbound$ diquarks, 
since the assumed bosonic character of the diquark favors the transition of 
free quarks at the Fermi surface to 
the bosonic ground state and ensures 
a global energy decrease, more specifically a lower value of the energy per 
baryon, 
for the appropiate choice of parameters. 
That is, since typical Fermi energies in a two flavor quark mixture is 
$\sim$ 300 MeV, the energy 
gained by condensing bosons and sending them to the ground state is in fact 
more relevant for stellar structure 
than the hypothetically gained energy of the pair gap. Moreover, even in the 
case of unbound diquarks 
a lower energy state is reached by the system as a whole. 
Therefore, although the parameter that mimics the pair gap may be not only 
negative but also zero or positive, 
energy is gained in all cases due to the assumed bosonic character of the 
superconducting pair. 
A slightly different parametrization used in \cite{HLP} produced different 
results. In particular the resulting stellar models were not so compact, a 
feature due to a larger stifness of that equation of state for all the 
parameter range. This gauges the sensitivity of the models to the 
assumptions on the basic physics. 

Finally, we must note that, strictly speaking,  quark-diquark matter is  
unstable against weak interactions producing strangeness. 
In this sense, the present study may be considered as a simplification analogous to pure 
neutron matter models of neutron stars. A more general study should include the possibility 
of strangeness production, the appeareance of leptons in the mixture, as well as 2SC and 
color flavor locking (CFL). We can expect that the opening of the weak 
channel will produce an even softer equation of state and so higher compactness of 
these stars. These improvements will be presented in future work.

\nonumsection{Acknowledgements}
\noindent G. Lugones acknowledges the Instituto Astron\'omico 
y Geofisico de S\~ao Paulo and the financial support received from the 
Funda\c c\~ao de Amparo \`a Pesquisa do Estado de 
S\~ao Paulo. J.E. Horvath wishes to acknowledge 
the CNPq Agency (Brazil) for partial financial support. 

\nonumsection{References}
 
\eject


\begin{figure} 
\fcaption{Equation of state of quark diquark matter in the quark mass density dependent model for different set 
of parameters $(C^{1/2}, m_{D0})$.  The curves are labeled with: a = $(160,-100)$, b = $(160,0)$,   c = $(200,-100)$,
d = $(160,210)$, e = $(185,100)$ and   f =  $(185,130)$.
For the sake of comparison we also include the ultrarelativistic EOS $P=E/3$ 
and the EOS presented in  Ref.[18]  (labeled as  HP).} 
\end{figure}

\begin{figure} 
\fcaption{The particle number densities of $u$ and $d$ quarks and diquarks for the parameters   $(160,200)$  (label A)
and   $(185,130)$ (label B) . At high baryon number densities
the number of quarks $u$ goes to zero and the number of quarks $d$ and diquarks are equal. We also include the case of 
pure quark d - diquark matter $n_B = n_d = n_D$ corresponding to any set of parameters in the region A' of Figure 3.} 
\end{figure}

\begin{figure} 
\fcaption{Different stability regions in the parameter space of the model. Inside regions A and $A^{'}$  
quark-diquark matter has a lower energy per baryon than a neutron being so absolutely stable. 
For parameters inside the  region $A^{'}$ there are no $u$ quarks  
in the mixture, and matter is always composed by quarks $d$ and diquarks.
Inside the region B below the full straight line the mixture has
a lower energy than  pure two flavor quark matter composed only by quarks $u$ and $d$.}
\end{figure}

\begin{figure} 
\fcaption{The mass radius relation for quark-diquark stars. The curve labeled with HP
corresponds to the EOS developed in \cite{horvath1}.
The other curves correspond  to the following 
set of the parameters $(C^{1/2}, m_{D0})$ : a = $(185,130)$,  b=$(185,100)$,  c=$(180,100)$,  d=$(160,210)$. 
Models overlap the range of the M-R measurements of Her X-1 and RX J1856-37.}
\end{figure}

\begin{figure} 
\fcaption{The same as the previous figure but for A= $(210,-100)$, B= $(200,-100)$, C= $(160,0)$,  D=$(160,-100)$.}
\end{figure}

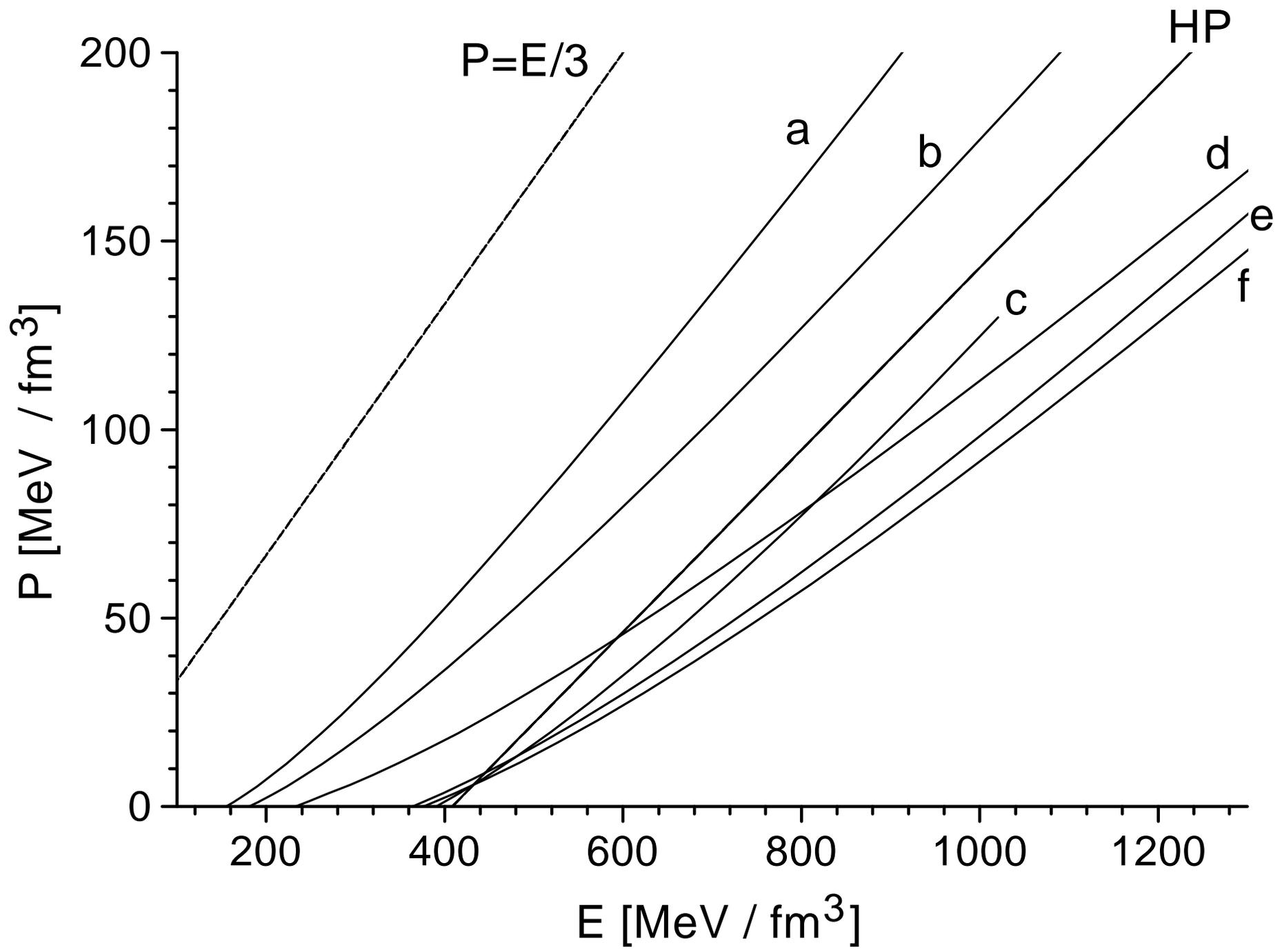
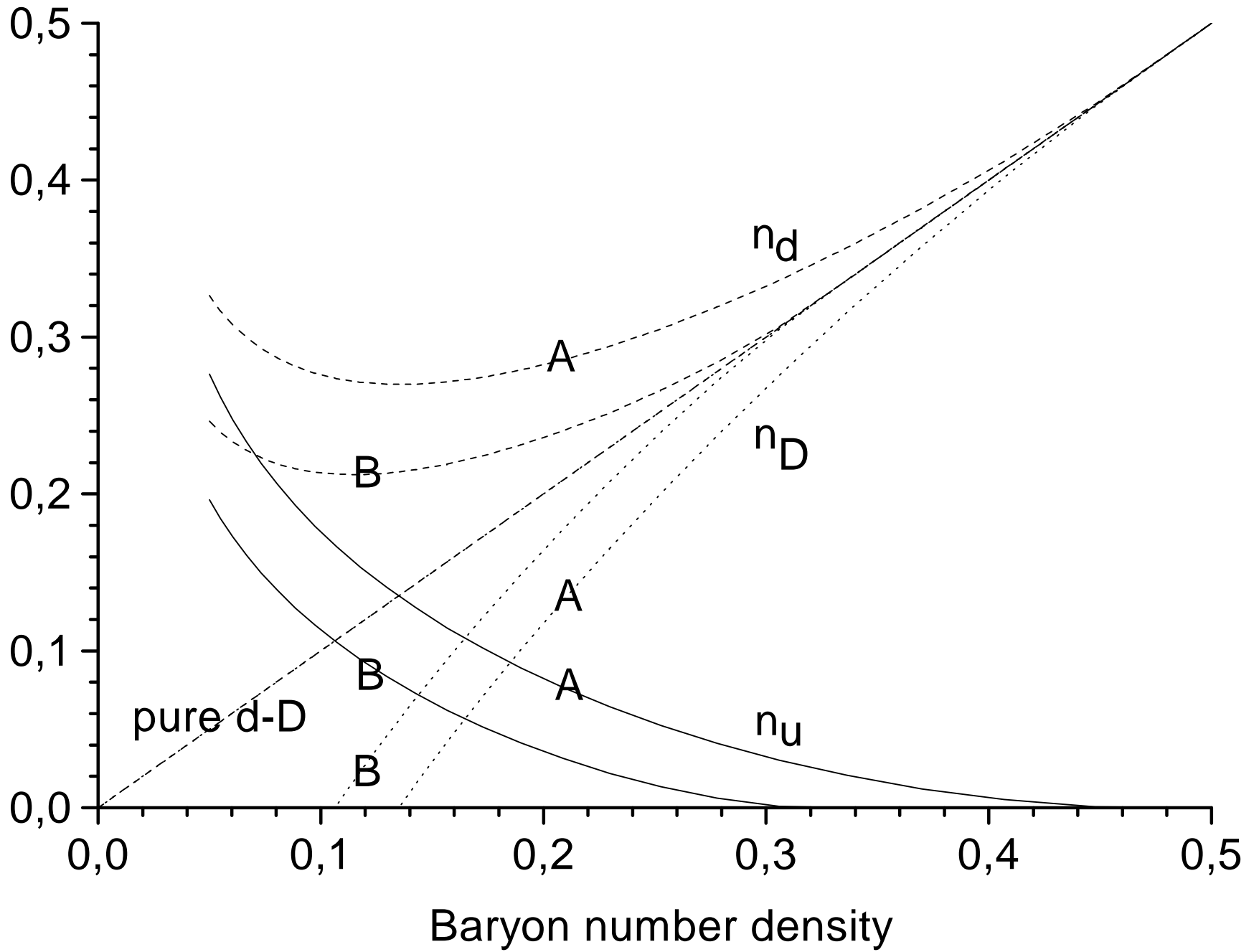
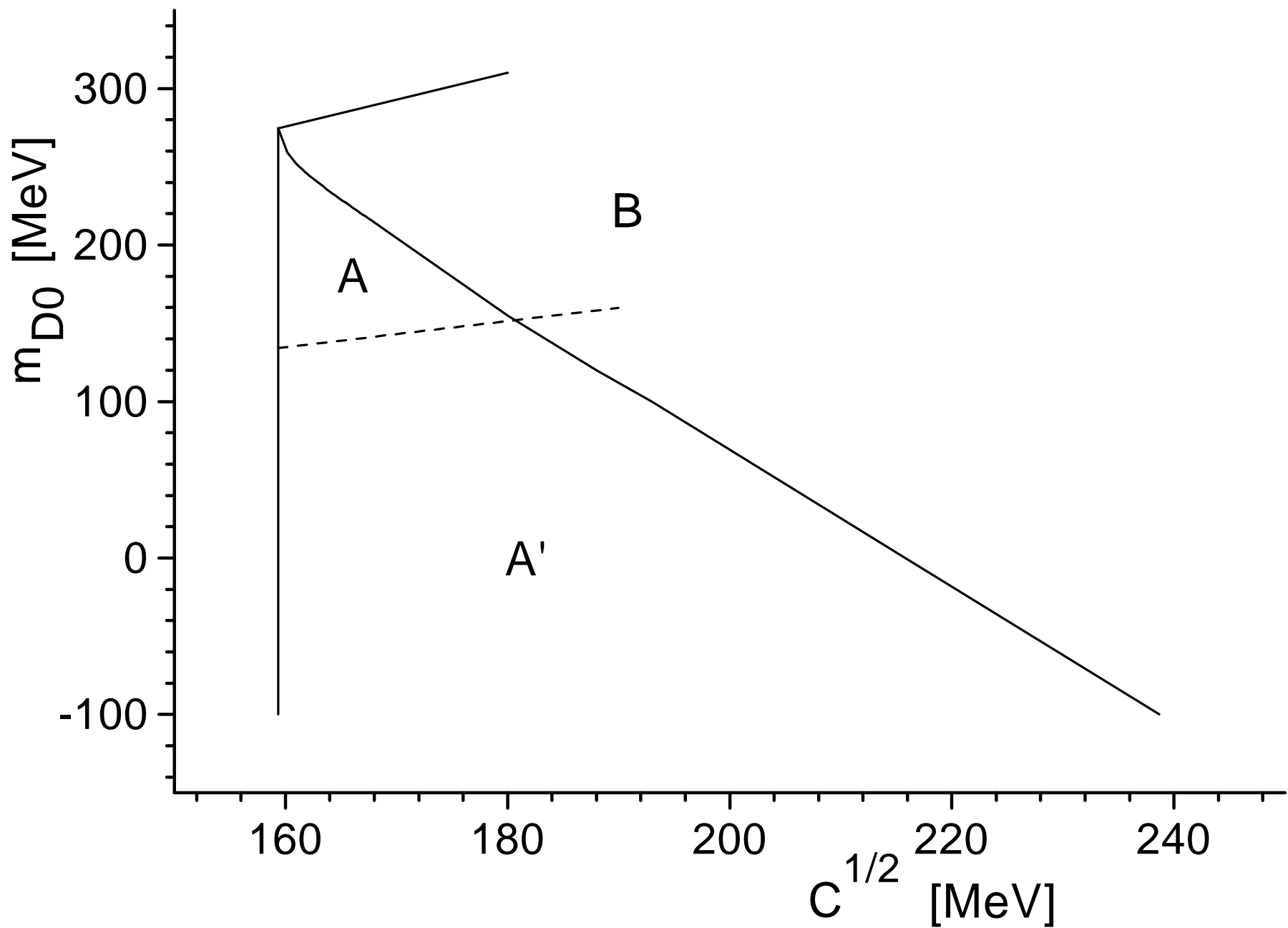
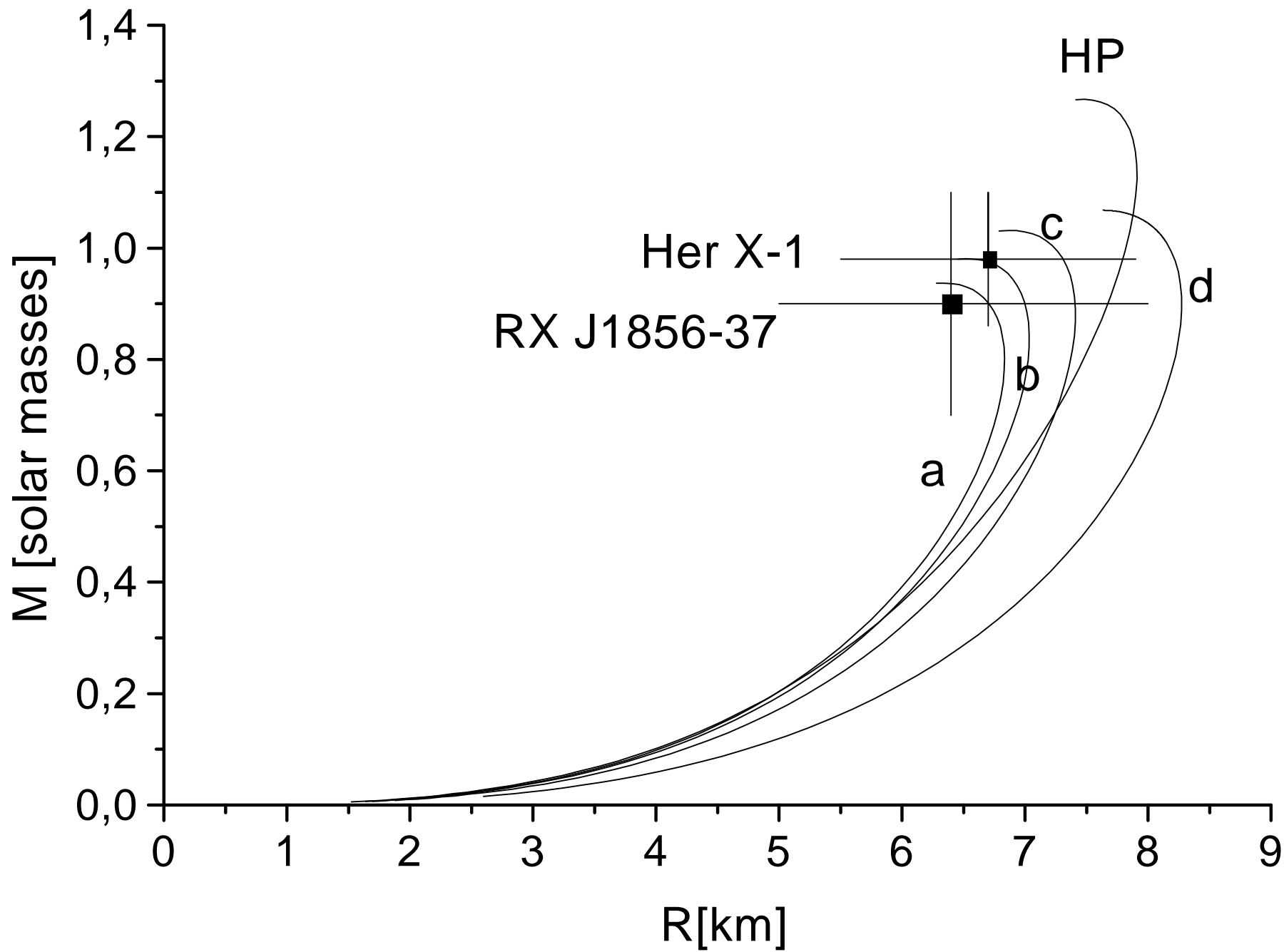
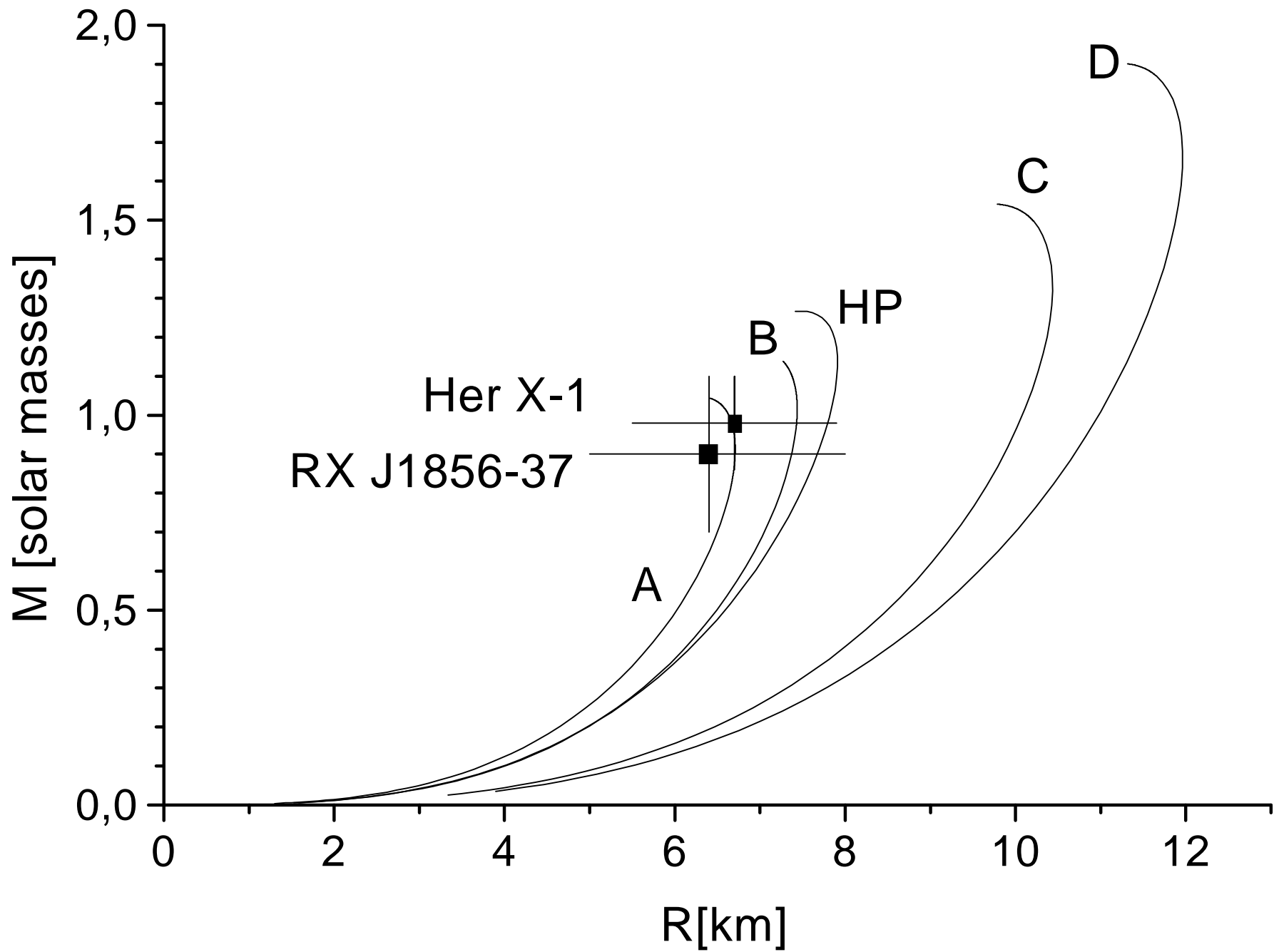
\end{document}